\documentclass[aps,prl,twocolumn,showpacs,nofootinbib,floatfix,superscriptaddress,preprintnumbers]{revtex4}
\usepackage{amsmath}
\usepackage{amssymb}
\usepackage{epsfig}
\usepackage{graphicx}
\DeclareMathAlphabet{\mathsc}{OT1}{cmr}{m}{sc}
\newcommand {\ignore}[1]{}

\def\10{$SO(10)$}
\def\21{SU(2) $\otimes$ U(1) }

\def\422{$SU(4) \otimes SU(2) \otimes SU(2)$}
\def\321{SU(3) $\otimes$ SU(2) $\otimes$ U(1)}
\def\gsim{\raise0.3ex\hbox{$\;>$\kern-0.75em\raise-1.1ex\hbox{$\sim\;$}}}
\def\lsim{\raise0.3ex\hbox{$\;<$\kern-0.75em\raise-1.1ex\hbox{$\sim\;$}}}

\def\lsim{\raise0.3ex\hbox{$\;<$\kern-0.75em\raise-1.1ex\hbox{$\sim\;$}}}
\def\gsim{\raise0.3ex\hbox{$\;>$\kern-0.75em\raise-1.1ex\hbox{$\sim\;$}}}
\def\vev#1{\left\langle #1\right\rangle}

\newcommand{\AddrAHEP}{%
  AHEP Group, Institut de F\'{\i}sica Corpuscular --
  C.S.I.C./Universitat de Val{\`e}ncia \\
  Edificio Institutos de Paterna, Apt 22085, E--46071 Valencia, Spain}\newcommand{\AddrNiigata}{%
 Graduate School of Science and Technology, Niigata University, Niigata 950-2181, Japan}

 \newcommand{\ba}{\begin{array}}
\newcommand{\ea}{\end{array}}
\relax

\let\vev\VEV
\usepackage{color}

\def\321{$SU(3)\times SU(2)\times U(1)$}

\begin{document}

\title{ Relating quarks and leptons without grand-unification}  %\date{\today}
\author{S. Morisi} \email{morisi@ific.uv.es} \affiliation{\AddrAHEP}
\author{E. Peinado} \email{epeinado@ific.uv.es} \affiliation{\AddrAHEP}
\author{Yusuke Shimizu} \email{shimizu@muse.sc.niigata-u.ac.jp} 
\affiliation{\AddrNiigata}
\author{J. W. F. Valle} \email{valle@ific.uv.es} \affiliation{\AddrAHEP}
\date{\today}
\begin{abstract}

 In combination with supersymmetry, flavor symmetry may relate quarks
  with leptons, even in the absence of a grand-unification group.
  We propose an \321 model where both supersymmetry and the assumed
  $A_4$ flavor symmetries are softly broken, reproducing well the
  observed fermion mass hierarchies and predicting: (i) a relation
  between down-type quarks and charged lepton masses, and (ii) a
  correlation between the Cabibbo angle in the quark sector, and the
  reactor angle $\theta_{13}$ characterizing CP violation in neutrino
  oscillations.
\end{abstract}

\pacs{
11.30.Hv       % Flavor symmetries
14.60.-z       % Leptons
14.60.Pq       % Neutrino mass and mixing
14.80.Cp       % Non-standard-model Higgs bosons
14.60.St       % Non-standard-model neutrinos, right-handed neutrinos, etc.
23.40.Bw     % Weak-interaction and lepton (including neutrino) aspects of decay
}

\maketitle

\section{Introduction}

Understanding the observed pattern of quark and lepton masses and
mixing~\cite{nakamura2010review,Schwetz:2008er} constitutes one of the
deepest challenges in particle physics.
Flavor symmetries provide a very useful approach towards reducing the
number of free parameters describing the fermion
sector~\cite{Ishimori:2010au}.
It has long been advocated that grand unification offers a suitable
framework to describe flavor. In what follows we will adopt the
alternative approach, assuming that flavor is implemented directly at the
\321 level. 
Typically this requires several $SU(2)$ doublet scalars in order to
break spontaneously the flavor symmetry so as to obtain an acceptable
structure for the masses and mixing matrices.
(One may alternatively introduce ``flavons'' instead of additional
Higgs doublets, but in this case one would have to give up
renormalizability).

In order to construct a ``realistic'' extension of the Standard Model
(SM) with flavor symmetry one needs a suitable alignment of the scalar
vacuum expectation values (vevs) in the
theory~\cite{ma:2001dn,babu:2002dz,zee:2005ut,altarelli:2005yp}.
There are several multi-doublet extensions of the SM with flavor in
the market, but renormalizable supersymmetric extensions of the SM
with a flavor symmetry are only a few~\cite{PhysRevD.71.056006},
usually because the existence of additional Higgs doublets spoils the
unification of the coupling constants.

Here we choose to renounce to this theoretical argument, noting that
gauge coupling unification may happen in multi-doublet schemes due to
other effects.
What we now present is a supersymmetric extension of the SM based on
the $A_4$ group where all the matter fields as well as the Higgs
doublets belong to the same $A_4$ representation, namely, the
triplet. This leads us to two theoretical predictions. The first a
mass relation
\begin{equation}
\label{eq:ours}
\frac{m_{\tau}}{\sqrt{m_em_\mu}}\approx\frac{m_b}{\sqrt{m_d m_s}}~,
\end{equation}
involving down-type quarks and charged lepton mass ratios. Such
relation can be obtained by a suitable combination of the three
Georgi-Jarlskog (GJ) mass relations~\cite{Georgi:1979df},
\begin{equation}\label{GJ}
\begin{array}{lll}
m_b=m_\tau,&m_s=1/3 m_\mu,&m_d=3 m_e,
\end{array}
\end{equation}
which arise within a particular ansatz for the SU(5) model and hold at
the unification scale.  In contrast to eq. (\ref{GJ}), our relation
requires no unification group and holds at the electroweak scale. It
would, in any case, be rather robust against renormalization effects
as it involves only mass ratios.

The second prediction obtained in our flavor model is a correlation
between the Cabibbo angle for the quarks and the so-called ``reactor
angle'' $\theta_{13}$ characterizing the strength of CP violation in
neutrino oscillations~\cite{nunokawa:2007qh,bandyopadhyay:2007kx}.
Within a reasonable approximation we find
\begin{equation}
\label{eq:12q-13l}
\lambda_C\approx \frac{1}{\sqrt{2}} \frac{m_\mu m_b}{m_\tau m_s}
\sqrt{\sin^22 \theta_{13}}-\sqrt{\frac{m_u}{m_c}}\,.
\end{equation}
which arises mainly from the down-type quark
sector~\cite{Gatto:1968ss} with a correction coming from the up
isospin diagonalization matrix. This is a very interesting relation,
discussed below in more detail.\\[-.5cm]

\section{The Model}
\label{sec:themodel}

Here we propose a supersymmetric model based on an $A_4$ flavor
symmetry realized in an \321 gauge framework.
The field representation content is given in
Table~\ref{tab:Multiplet1}. Note that all quarks and leptons transform
as $A_4$ triplets.  Similarly the Higgs superfields with opposite
hypercharge characteristic of the MSSM are now upgraded into two sets,
also transforming as $A_4$ triplets. Note that since all matter fields
transform in the same way under the flavor symmetry one may in
principle embed the model into a grand-unified scheme. However, given
the large number of scalar doublets, gauge coupling unification must
proceed differently, see, for example, Ref.~\cite{Munoz:2001yj}.
\begin{table}[h!]
\begin{center}
  \begin{tabular}{|l||lllll||ll|}
\hline
fields & $\hat{L}$ & $\hat{E}^c$ & $\hat{Q}$& $\hat{U}^c$ & $\hat{D}^c$ & $\hat{H}^u $ & $\hat{H}^{d}$  \\
\hline
$SU(2)_L$ & 2&1&2& 1& 1&2&2 \\
$A_4$  &  3& 3 &3  & 3 &3&3&3 \\
\hline
\end{tabular}
\caption{Basic multiplet assignments of the model}
\label{tab:Multiplet1}
\end{center}
\end{table}

The most general renormalizable Yukawa Lagrangian for the charged
fermions in the model is~\cite{Morisi:2009sc}
\begin{equation}\label{y}
L_{\text{Yuk}}
= y^l_{ijk}\hat{L}_i\hat{H}^d_{j}\hat{E}^c_k
+ y^d_{ijk}\hat{Q}_i\hat{H}^d_{j}\hat{D}^c_k
+ y^u_{ijk}\hat{Q}_i\hat{H}^u_{j}\hat{U}^c_k~,
\end{equation}
where $y_{ijk}^{u,d,l}$ are $A_4$-tensors, assumed real at this
stage.

The Higgs scalar potential invariant under $A_4$ is
\begin{equation}\label{eq:V}
\begin{array}{lll}
V&=&
 (|\mu |^2+m_{H_u}^2)(|H_1^u|^2+|H_2^u|^2+|H_3^u|^2)\\
&+&(|\mu |^2+m_{H_d}^2)(|H_1^d|^2+|H_2^d|^2+|H_3^d|^2) \\
&-&[b(H_1^uH_1^d+H_2^uH_2^d+H_3^uH_3^d)+\text{c.c.}] \\
&+&\frac{1}{8}(g^2+{g^\prime }^2)(|H_1^u|^2+|H_2^u|^2+ |H_3^u|^2+\\
&&\qquad\qquad -|H_1^d|^2-|H_2^d|^2-|H_3^d|^2)^2~. \\
\end{array}
\end{equation}

Assuming that the Higgs doublet scalars take real vevs $\vev{
  H_i^{u,d}} = v_i^{u,d}$ one can show that the minimization of the
potential $V$ gives as possible local minima the alignments
$\vev{{H^{0}}^{u,d}} \sim (1,0,0)$ and $(1,1,1)$.  Only the first is
viable and we verify that minimization leads to this solution within a
wide region of parameters.  By adding $A_4$ soft breaking terms to the
$A_4$-invariant scalar potential in eq.~(\ref{eq:V})
\begin{equation}\label{align0}
\begin{array}{lll}
V_{soft}&=&\sum_{ij}\left(\mu^u_{ij}H_i^{u*}H_j^u+\mu^d_{ij}H_i^{d*}H_j^d\right)+\sum_{ij}b_{ij}H_i^{d}H_j^u, \nonumber
\end{array}
\end{equation}
one finds that
\begin{eqnarray}
\label{eq:minima}
\vev{H^u}=(v^u,\varepsilon_1^u,\varepsilon_2^u),\quad
\vev{H^d}=(v^d,\varepsilon_1^d,\varepsilon_2^d)~,
\end{eqnarray}
where $\varepsilon_{1,2}^u\ll v^u$ and $\varepsilon_{1,2}^d\ll v^d.$\\[-.5cm]

\subsection{Charged fermions}
\label{sec:charged-fermions}

By using $A_4$ product rules it is straightforward to show that the
charged fermion mass matrix takes the following universal
structure~\cite{Morisi:2009sc}
\begin{equation}
M_{f}=\left(
\begin{array}{ccc}
0 & y_1^f \vev{ H_3^f } & y_2^f \vev{ H_2^f } \\
y_2^f \vev{ H_3^f } & 0 & y_1^f \vev{ H_1^f } \\
y_1^f \vev{ H_2^f } & y_2^f \vev{  H_1^f } & 0
\end{array}
\right),
\label{eq:me}
\end{equation}
where $f$ denotes any charged lepton, up- or down-type quarks.  Note
that, in addition to the ``texture'' zeros in the diagonal, one has
additional relations among the parameters.  This may be seen
explicitly by rewriting eq.~(\ref{eq:me}) as
\begin{equation}
M_{f}=
\left(
\begin{array}{ccc}
0 & a^f \alpha^f  & b^f  \\
b^f\alpha^f  & 0 & a^f r^f \\
a^f  & b^f r^f & 0
\end{array}
\right),
\label{Mf}
\end{equation}
where $a^f=y_1^f\varepsilon_1^f$, $b^f=y_2^f\varepsilon_1^f$, with
$y_{1,2}^f$ denoting the only two couplings arising from the
$A_4$-tensor in eq.~(\ref{y}), $r^f=v^f/\varepsilon_1^f$ and
$\alpha^f=\varepsilon_2^f/\varepsilon_1^f$.  Thanks to the fact that
the same Higgs doublet $H^d$ couples to the lepton and to the
down-type quarks one has, in addition, the following relations
\begin{equation}\label{rel}
r^l=r^d,\qquad \alpha^l=\alpha^d,
\end{equation}
involving down-type quarks and charged leptons.

\begin{center} 
\begin{figure}[h!] 
\includegraphics[width=7cm]{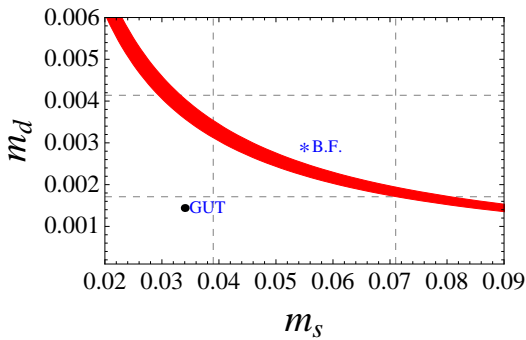}
\caption{The shaded band gives our prediction for the down-strange
  quark masses at the $M_z$ scale, eq.~(\ref{massrel2}), vertical and
  horizontal lines are the 1$\sigma$ experimental ranges from
  ref.~\cite{Xing:2007fb}. }
\label{figmass}
\end{figure}
\end{center}

It is straightforward to obtain analytical expressions for $a^f$,
$b^f$ and $r^f$ from eq.~(\ref{Mf}) in terms of the charged fermion
masses and $\alpha^f$,
\begin{equation}
\label{eq:rel2}
\frac{r^f}{\sqrt{\alpha^f}}\approx \frac{m_{3}^f}{\sqrt{m_{1}^fm_{2}^f}},\quad
a^f\approx\frac{m_{2}^f}{m_{3}^f}\frac{\sqrt{m_{1}^f{m_{2}^f}}}{\sqrt{\alpha^f}},\quad
b^f\approx \frac{\sqrt{m_{1}^f{m_{2}^f}}}{\sqrt{\alpha^f}}.\quad
\end{equation}
From eq.~(\ref{rel}) and eq.~(\ref{eq:rel2}) it follows that 
\[
 \frac{m_{\tau}}{\sqrt{m_{e}\, m_{\mu}}}\approx  \frac{m_{b}}{\sqrt{m_{d}\, m_{s}}},
\]
a formula relating quark and lepton mass ratios (to a very good
approximation this formula also holds for complex Yukawa couplings).
This relation is a strict prediction of our model, and appears in a
way similar to the celebrated SU(5) mass relation, despite the fact
that we have not assumed any unified group, but just the \321 gauge
structure. It allows us to compute the down quark mass in terms of the
charged fermion masses and the $s$ and $b$ quarks, as
\begin{equation}\label{massrel2}
 m_d\approx m_{e}  \frac{m_{\mu}}{ m_{s}}\left(\frac{m_{b}}{m_{\tau}}\right)^2.
\end{equation}
This mass formula predicts the down quark mass at the scale of the $Z$
boson mass, to lie in the region
\begin{equation}
\begin{array}{c}
1.71~ MeV <m_d^{th}<3.35 ~MeV \\ 1.71~ MeV <m_d< 4.14 ~MeV~,
\end{array}\end{equation}
at 1$\sigma$~\cite{Xing:2007fb}. This is illustrated in
Fig.~\ref{figmass} where, to guide the eye, we have also included the
1$\sigma$ experimental ranges from reference~\cite{Xing:2007fb}, as
well as the best fit point and the GJ prediction.\\[-.5cm]
Note also that, thanks to supersymmetry, we obtain a relation only
among the charged lepton and down-type quark mass ratios, avoiding the
unwanted relation found by Wilczek and Zee in
Ref.~\cite{Wilczek:1978xi}.\\[-.5cm]

\subsection{Neutrinos}
\label{sec:neutrinos}

To the renormalizable model we have so far we now add an effective
dimension-five $A_4$-preserving lepton-number violating operator
\begin{equation}\label{nudim5}
\mathcal{L}_{5d}=
\frac{f_{ijlm}}{\Lambda} \hat{L}_i \hat{L}_j \hat{H}^u_l \hat{H}^u_m ~,
\end{equation}
where the $A_4$-tensor $f_{ijlm}$ takes into account all the possible
contractions of the product of four $A_4$ triplets~\footnote{ Specific
  realizations of $\mathcal{L}_{5d}$ within various seesaw
  schemes~\cite{Valle:2006vb} can, of course, be envisaged.}.

Neutrino masses are induced after electroweak symmetry breaking from
the operator in eq.\,(\ref{nudim5}). In order to determine the flavor
structure of the resulting mass matrix we take the limit where the vev
hierarchy $\vev{H_1^u} \gg \vev{H_2^u}, \vev{H_3^u}$ holds, leading
to~\cite{Morisi:2009sc}
\begin{equation}
M_\nu=\left(\begin{array}{ccc}
    x {r^u}^2  & \kappa r^u  & \kappa r^u \alpha^u\\
    \kappa r^u  & y {r^u}^2  & 0\\
    \kappa r^u \alpha^u& 0 & z {r^u}^2
\end{array}
\right),
\label{mnu20}
\end{equation}
where $x,\,y,\,z$ and $\kappa$ are coupling constants, while $r^u$ and
$\alpha^u$ already been introduced above in the up quark sector.

The best fit of neutrino oscillation data~\cite{Schwetz:2008er} yields
maximally mixed $\mu$ and $\tau$ neutrinos. This is possible, in the
basis where charged lepton is diagonal, if and only if the
light-neutrino mass matrix is approximately $\mu-\tau$ invariant. In
turn this holds true if $y \approx z$ and $\alpha^u \approx
1$~\cite{Morisi:2009sc}\footnote{The charged lepton mass matrix is
  mainly diagonalized by a rotation in the 12 plane.}. When
$\alpha^u<1$ the ``atmospheric angle'' deviates from the
maximality. We have verified that for $\alpha^u\gtrsim 0.5$
the atmospheric angle is within its 3~$\sigma$ allowed range.\\[-.5cm]

\section{Relating the Cabibbo angle to $\theta_{13}$}
\label{sec:relating}

In the CP conserving limit we have taken so far we have in total 14
free parameters to describe the fermion sector: six $a^f$ and $b^f$
parameters (three for each charged fermion-type), plus four $r^f$ and
$\alpha^f$ (here only down-type are counted, in view of
eq.~(\ref{rel})), plus four parameters describing the neutrino mass
induced by the dimension-5 operator: $x, y, z, \kappa$.
These parameters describe 18 observables, which may be taken as the 9
charged fermion masses, the two neutrino squared mass differences
describing neutrino oscillations, the three neutrino mixing angles,
the neutrinoless double beta decay effective mass parameter, the
Cabbibo angle, in addition to $V_{ub}$ and $V_{cb}$. Hence we have
four relations.

The first of these we have already seen, namely the mass relation in
eq.~(\ref{eq:ours}) and Fig.~\ref{figmass}.
The second is a quark-lepton mixing angle relation concerning the
Cabibbo angle $\lambda_C$ and the ``reactor angle'' $\theta_{13}$
describing neutrino oscillations.  To derive it note first that the
matrix in eq.\,(\ref{Mf}) is diagonalized on the left by a rotation in
the 12 plane, namely
\begin{equation}\label{R12}
\sin\theta^f_{12}\approx\sqrt{\frac{m_1^f}{m_2^f}} \frac{1}{\sqrt{\alpha^f}}.
\end{equation}

\begin{center} 
\begin{figure}[h!] 
\vglue -1cm
\includegraphics[width=7cm]{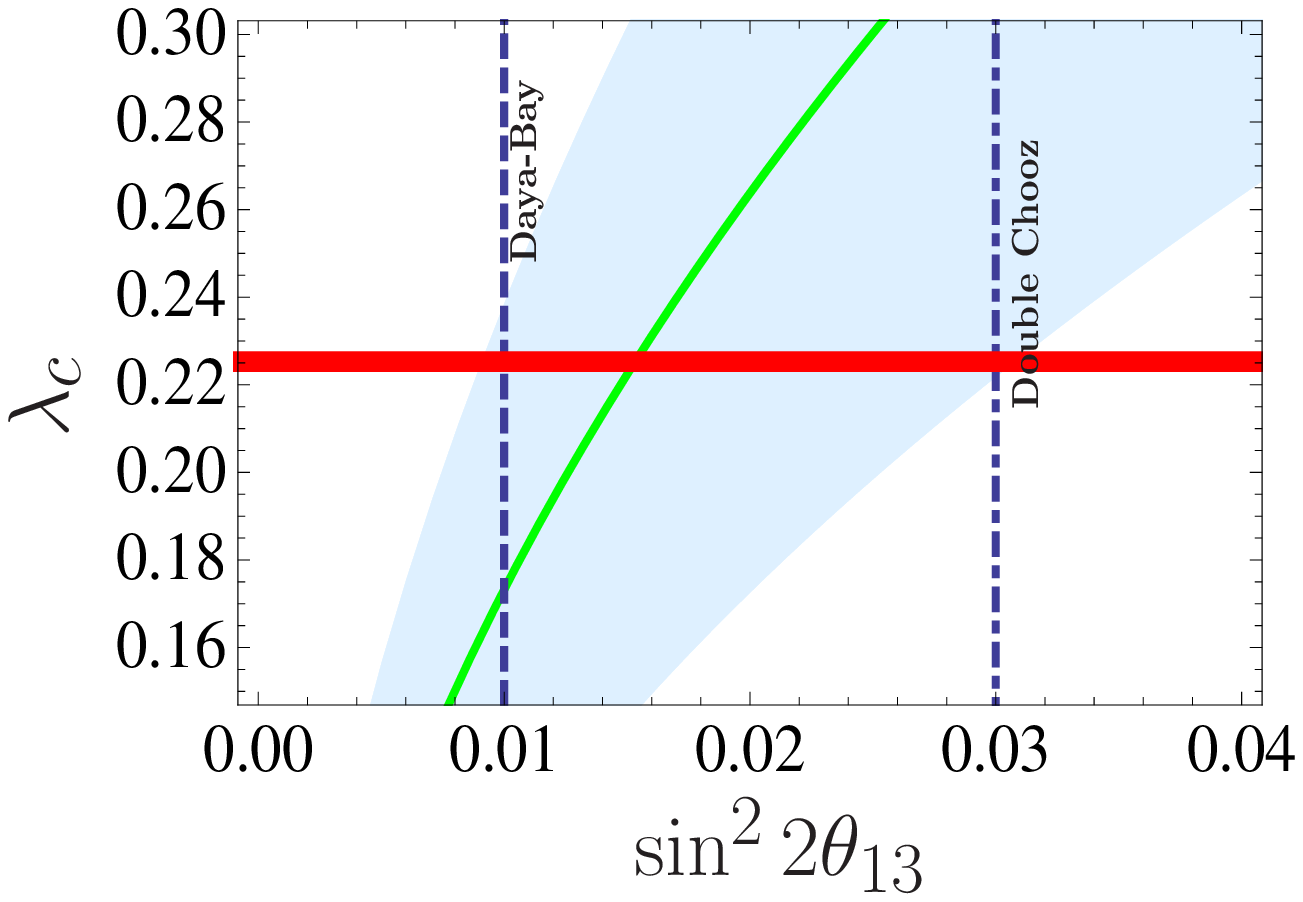}
\caption{The shaded band gives our predicted 1~$\sigma$ correlation
  between the Cabibbo angle and the reactor angle, as above. Vertical
  lines give the expected sensitivities on
  $\theta_{13}$~\cite{Ardellier:2006mn,Guo:2007ug}.}
\label{fig2}
\end{figure}
\end{center}

In order to give an analytical expression for the relation between
Cabbibo and reactor angles, we neglect mixing of the third family of
quarks and go in the limit where our neutrino mass matrix, eq. (15) is
$\mu-\tau$ invariant, that is $\alpha^u=1$ and $y=z$. In this
approximation, the reactor mixing angle is given by
\begin{equation}
\sin \theta_{13}=\frac{1}{\sqrt{2}} \sin \theta_{12}^l=\frac{1}{\sqrt{2}}\sqrt{\frac{m_e}{m_\mu}}\frac{1}{\sqrt{\alpha^l}},
\label{t13-1}
\end{equation}
using our mass relation in eq.~(\ref{eq:ours}) one finds that the
Cabbibo angle may be written as
\begin{equation}\label{tc-1}
\lambda_C=\frac{m_b}{m_s}\frac{\sqrt{m_e m_\mu}}{m_\tau}\frac{1}{\sqrt{\alpha^d}}-\sqrt{\frac{m_u}{m_c}}.
\end{equation}
Comparing eq. (\ref{t13-1}) with eq. (\ref{tc-1}) leads immediately to
equation (\ref{eq:12q-13l}). 
In order to display this prediction graphically we take the quark
masses at 1~$\sigma$, obtaining the curved band shown in
fig.~\ref{fig2}.
The narrow horizontal band indicates current determination of the
Cabbibo angle, while the two vertical dashed lines represent the
expected sensitivities of the Double-Chooz~\cite{Ardellier:2006mn} and
Daya-Bay~\cite{Guo:2007ug} experiments on the ``reactor mixing angle''
$\theta_{13}$. The curved line corresponds to the analytical
approximation for the best fit value of the quark masses in
eq.~(\ref{eq:12q-13l}). Clearly the width of the curved band
characterizing our prediction is dominated by quark mass determination
uncertainties.\\[-.5cm]

Finally note that mixing parameters of the third family of quarks $
U_{13}^q\approx \frac{m_2^q}{m_3^q}\frac{\sqrt{m_1^q
    m_2^q}}{m_3^q}\frac{1}{\sqrt{\alpha^q}}$ and $U_{23}^q\approx
\frac{m_1^q (m_2^q)^2}{(m_3^q)^3}\frac{1}{\alpha^q}$ ($q=u,d$)
are negligible, and can not account for the measured values of
$V_{ub}$ and $V_{cb}$. The predicted values obtained for these are too
small so that in its simplest presentation described above our model
can not describe the CP violation found in the decays of neutral
kaons.  However there is a simple solution which maintains the good
predictions described above, namely, adding colored vector-like
$SU(2)_L$ singlet states. In this case acceptable values for $V_{ub}$
and $V_{cb}$, leading to adequate CP violation can arise solely from
non-unitarity effects
of the quark mixing matrix.\\[-.5cm]

\section{Outlook}
\label{conc}

We proposed a supersymmetric extension of the standard model with an
$A_4$ flavor symmetry, where all matter fields in the model transform
as triplets of the flavor group.  Charged fermion masses arise from
renormalizable Yukawa couplings while neutrino masses are treated in
an effective way. 
The scheme illustrates how, in combination with supersymmetry, flavor
symmetry may relate quarks with leptons, even in the absence of a
grand-unification group.
Two good predictions emerge: (i) a relation between down-type quarks
and charged lepton masses, and (ii) a correlation between the Cabibbo
angle in the quark sector, and the reactor angle $\theta_{13}$
characterizing CP violation in neutrino oscillations, which lies
within the sensitivities of upcoming experiments.

Although the predicted values for the other mixing parameters $V_{uc}$
and $V_{cb}$ of the Cabibbo-Kobayashi-Maskawa matrix are too small, we
mentioned a simple way to circumvent this, making the scheme fully
realistic. 

Finally note that, with few exceptions such as those in
Refs.~\cite{King:2003rf,Dermisek:2005ij}, grand-unified flavor models
are not more predictive than the novel idea proposed here and
illustrate through this simple scheme.
As it stands the model fits well with the idea that gauge coupling
unification may be an effect of the presence of extra dimensions
rather than of grand-unified interactions~\cite{Munoz:2001yj}.
Notwithstanding, we wish to stress that our model is manifestfly
embeddable into a standard Grand-Unified scenario, which would result
in further predictive power.  A detailed study of this particular
model lies outside the scope of this letter and will be taken up
elsewhere.

We thank M. Hirsch and M.Tanimoto for useful discussions.  This work
was supported by the Spanish MICINN under grants FPA2008-00319/FPA and
MULTIDARK CSD2009-00064, by Prometeo/2009/091 (Generalitat
valenciana), by the EU Network grant UNILHC
PITN-GA-2009-237920. S. M. is supported by a Juan de la Cierva
contract, E. P. is supported by CONACyT (Mexico) and Y.S. by the Japan
Society of Promotion of Science, Grand-in-Aid for Scientific Research,
No.22.3014.

\end{document}